\documentclass[preprint]{aastex}

\shorttitle{NGC 5128 GCS}
\shortauthors{Harris et al.}

\begin{document}

\title{A Two-Phase Chemical Enrichment Model \\
for the Milky Way Globular Cluster System}

\author{Marcel L.~VanDalfsen and William E.~Harris}
\affil{Department of Physics and Astronomy, McMaster University, Hamilton ON L8S 4M1}
\email{vandalfs@physics.mcmaster.ca, harris@physics.mcmaster.ca}

\begin{abstract}

Many globular cluster systems have a distinct
bimodal metallicity distribution function (MDF) which has strikingly
similar features in many large galaxies of all types.  
By using the Milky Way cluster system as a typical example, we show
that bimodal MDFs can be very well matched with a double
``accreting-box'' chemical enrichment model in 
which both the halo (metal-poor) and
bulge (metal-rich) clusters form during an early phase of gas
inflow simultaneously with star formation.  However, 
differences in effective yield between the two phases are not enough 
by themselves to reproduce the observed MDF shape:  gas infall is required for
both phases, and either the initial gas or the infalling gas must
have very different metallicities in the two separate phases.

\end{abstract}


\clearpage

\section{Introduction}

The old globular clusters found in all large galaxies
have presented long and persistent challenges to our understanding of
the early formation processes of galaxy halos and bulges.  A feature of 
globular cluster systems (GCSs) which has drawn considerable attention
is their metallicity distribution (MDF), which consistently
takes on a roughly bimodal form in many large elliptical and large spiral
galaxies \citep[e.g.][among many others]{whi95,geb99,zep95,gei96,for97,
nei99,kun01,lar01,lfb01,bea03,har03}.  
Most clusters fall either in the ``metal-poor''
mode centered at [Fe/H] $\simeq -1.5$ or ``metal-rich'' mode centered
near [Fe/H] $\simeq -0.5$, with each mode having an intrinsic dispersion
$\sigma$[Fe/H] $\simeq 0.3$ dex.
Furthermore, the mean [Fe/H] values and 
dispersions of each mode are surprisingly similar from one galaxy to the next
despite the vast range of galaxy types 
\citep[e.g.][]{for97,h01,for01}.\footnote{To date, most of the MDF data for
the many galaxies in the studies cited above are from broadband photometric indices
such as $(V-I)$, which are intrinsically not very sensitive to metallicity.
In addition, the two standard ``modes'' have Gaussian 
dispersions large enough to allow a significant degree of intrinsic overlap,
so that if they are blurred out further by low-precision data, the validity
of a bimodal fit becomes debatable.    
For examples of MDFs where
the two modes may not have clearly separated peaks, see, e.g., 
NGC 524, 1399, or 4473 in \citet{lar01}.  It is, however,
important to note that when data are taken with
higher-precision photometry or spectroscopy, bimodal distributions tend
to emerge more clearly.  That is, the galaxies with the best available
data (such as M87, NGC 4472, NGC 1404, NGC 4594, NGC 5128, or the Milky Way itself;
see the references cited above) are very well matched by a bimodal MDF
in which the two modes display clear separation.  It is these 
``best case'' examples which have driven the bimodality paradigm in the
literature and which our present discussion is aimed at.}

The Milky Way GCS is no exception to this general trend.  Because it is
a completely uncontaminated sample of clusters with well established
metallicities, it provides what is still perhaps the best classic example 
of a bimodal MDF (see Figure 1).  The shape of the MDF 
and its close connection with the cluster kinematics was first
clearly established by \citet{zin85} and has held up through many subsequent
analyses \citep[see, e.g.][for a review]{h01}.
The MDF, in its usual plotted form as number per unit [Fe/H] (left panel of
Figure 1), is often approximated by a pair of Gaussian functions 
(cf. the references cited above).
These numerical fits have no physical significance (and have not usually been
claimed to), and give little insight into the evolutionary history
of the system except for the strong hint that a distinct two-phase
formation history is required.  

In this brief paper, we take one step further into interpreting
these bimodal MDFs 
using some simple and physically plausible assumptions.
Standard first-order chemical evolution models, such as the
one-zone closed-box or ``simple model'' 
\citep{pag75,bin98} work reasonably well at
matching the broad, unimodal MDFs of the halo field-star populations 
in the Milky Way \citep[e.g.][]{pag75,har76,rya91,pra03} 
and in M31 \citep{dur01} with appropriate choices of
the effective nucleosynthetic yield $y_{\rm eff}$.
\footnote{For a normal IMF and a closed-box system, 
the nucleosynthetic yield is typically $y_0 \simeq 0.008$.
In ``leaky-box'' models, the value of $y_{\rm eff}$ can be much smaller and 
is strongly driven by the fraction of gas 
ejected from the zone by supernovae and stellar winds 
at each round of star formation; see Hartwick (1976).}  
Extensions to the simple model to incorporate
some smoothly varying gas inflow (the ``accreting-box'' model) 
have recently been used to model the unimodal, metal-rich 
MDF of the halo stars in NGC 5128
\citep{har02}.  An early, brief phase of primordial gas inflow 
has even been applied to explain the relatively
small number of the {\sl very} lowest-metallicity stars ([Fe/H] $< -3$)
in the Milky Way halo \citep{pra03}.
Accreting-box models can be thought of as 
approximate versions of more extensive
hierarchical-merging models of galaxy formation, in which a large
population of small pregalactic gas clouds merges successively into bigger and
bigger pieces while star formation within the clouds goes on  
simultaneously \citep[see, e.g.][for an application of the GALFORM code
to the case of NGC 5128]{bea03}.

By contrast, the distinctly bimodal MDFs of globular cluster systems 
do not easily fit into the continuous
sequence of starforming events in
normal hierarchical-merging or accreting-box models.
Some new feature of the history needs to be invoked, either
to boost the formation of metal-poor clusters at early times,
or inhibit the slightly later formation of intermediate-metallicity
clusters.  \citet{har02} and \cite{bea02} suggest that the
metal-poor clusters may have formed preferentially early in
the first rounds of star formation, but that their formation
was then truncated while
field-star formation continued onward.  In this picture,
the metal-rich clusters would have formed a bit later, during
the major starbursts accompanying the merging of the last, biggest
gas clouds.  \citet{san02}, for example, suggests that the metal-poor cluster
formation epoch was truncated at the reionization era.

An alternate approach \citep{cot98,cot02}
is to assume that the metal-poor clusters all 
formed separately in dwarf galaxies
and were dissipationlessly accreted later by the giants.  In this case
the bimodality of the combined MDF is a stochastic accident which results
only for a restricted mass range and mass spectrum of the accreted satellites.
\citet{ash92} proposed instead that the metal-poor clusters formed
{\sl in situ} while the metal-richer clusters formed later from gas
brought in by major mergers (though this model was primarily intended
to explain E-galaxy formation rather than big spirals).  Both
of these routes can produce bimodal MDFs, though many other interpretive
problems exist \citep[see, e.g.,][for extensive
discussions]{for97,h01,hh01}.

\section{A Two-Stage Chemical Enrichment Model}

We now use the Milky Way GCS to explore just one particular aspect
of the issues introduced above:  namely, how a simple ``accreting-box''
chemical evolution model can be adapted to produce a bimodal MDF.
Such a model might also allow us to gain insight into important
questions about the MDF:  Can we usefully constrain the initial gas
abundance of either major stage, or the abundance of the infalling gas?
Are the effective yield rates of each phase well determined once we
drop the closed-box restriction?  What about the time period over which
the infall was important?

The raw MDF for the Milky Way GCS is shown in Figure 1 in two versions:
number per unit [Fe/H], and number per unit heavy-element abundance $Z$.
We have used log $(Z/Z_{\odot}) \simeq [Fe/H] + 0.3$ for the standard 
conversion to total heavy-element abundance \citep[e.g.][]{she01}.
The linear form of the MDF, $dn/dZ$, connects 
readily with simple enrichment models, but in either version the observed
histogram is bimodal, with a primary peak at $Z \simeq 0.06 Z_{\odot}$
or [Fe/H] $=-1.5$ (the halo clusters) and a secondary peak 
at $Z \simeq 0.5 Z_{\odot}$ or [Fe/H] $=-0.5$ (the bulge clusters).  

For our extension of the simple model we follow the formalism
of \citet{har02} and explicitly assume that the GCS forms
in two distinct, separate stages.
Measurements of GC ages versus metallicity
\citep[e.g.][]{van00} suggest that these halo and bulge
phases may have been separated by about 2 Gy in time but
with some overlap between them.

Within each of the two phases, we assume that the GCs are built during
a large number of individual star-forming events, starting from a box
of gas with initial heavy-element abundance $Z_0$ and initial mass $M_0$.
In each discrete timestep $\delta t$ 
a small fraction $e$ of the gas reservoir $M_g$
forms into new stars.  A fraction $\alpha$ stays
locked up in dead stellar remnants, and $(1-\alpha)$ is returned
to the reservoir of gas in the box, so that the net increase
in stellar mass is $\delta M_s = \alpha e M_g$.  We assume ``prompt mixing'', i.e.
the ejected gas mixes with the surrounding gas
before the next star-forming step.  The effective yield $y$ by convention
is the amount of heavy elements released into the gas reservoir per unit
stellar remnant mass.  If the box simply uses up its initial supply
of gas, then after many timesteps the resulting MDF of stars formed in
the box has a characteristic exponential shape
$dn/dZ \sim e^{-(Z-Z_0)/y}$ \citep{pag75,bin98}.  
However, no such single-phase model will match Fig.~1(b).  

To add new gas continuously falling in to the system, we now assume
that an amount $M_{\rm add}$, with heavy-element abundance
$Z_{\rm add}$, comes in to the box after each timestep and
mixes with both the remaining initial gas and the gas ejected
from stars.  Under these conditions the net change of gas mass in
heavy elements after each timestep is \citep{har02}
\begin{equation}
\delta M_Z 
=  -Z \delta M_s \,
 + \, (1-Z)y \delta M_s \,
 + \, Z_{\rm add} M_{\rm add},
\end{equation}
where $Z \delta M_s$ is the mass of heavy elements removed
from the reservoir due to forming stars, 
$(1-Z)y \delta M_s$ is the amount added back as
ejected gas from SNe winds and evolved stars,
$Z_{\rm add} M_{\rm add}$ is the mass added by 
the new gas infalling into the reservoir,
and $\delta M_s  =  \alpha e M_g$.
For both stages (``halo'' and ``bulge'') we fix the numerical conversion
efficiency at $e=0.05$ and the ejected gas fraction at $(1-\alpha) = 0.2$
\citep[see][]{har02}; and
we then numerically integrate Eq.(1) over many $\delta t$ steps until the 
gas supply is depleted.  The metallicity of the ambient gas steadily climbs 
upward while gas inflow and star formation are both going on.  
The continual
accretion of gas into the system, combined with the recycling 
of enriched gas from each round of star formation, 
allows large numbers of stars
to be built in at intermediate and moderately high
metallicities, resulting in an MDF with a very different shape from
the closed-box model.

Following other authors, we use a gas infall rate which decreases smoothly
with e-folding time 
$\tau$, $dM_{\rm add}/dt = M_{\rm add}(0) e^{-t/\tau} $
\citep[see also][]{lac83,chi97,har02,pra03}.  Thus in the model as a whole,
the free parameters are the effective yield $y$; the initial abundance
$Z_0$; the abundance $Z_{\rm add}$; the initial amount $M_{\rm add}(0)$ of
infalling gas expressed in units of $M_0$; and the infall 
decay time $\tau$ expressed in units of the timestep $\delta t$.
Since each formation stage has its own set of
five parameters, the total two-stage model has a total of 10 parameters
to specify, plus
an 11th one which is the relative number of metal-rich to metal-poor clusters
(equivalent to the total amount of gas used in each stage).
At the end of the integration when the gas supply is nearly exhausted, 
we finally count up the numbers of stars 
$dn/dZ$ in the model as a function of abundance $Z$, for comparison with the observations.

In principle, we could solve for the entire set of parameters
by running a large grid of models and calculating goodness of fit
for each one.  
However, not all the parameters have important effects on the final MDF
shape, and some are mutually correlated.  
Here we explore just four different combinations of $Z_0$ and $Z_{add}$ which
have some physical motivation:

\noindent (a) $Z_0 = 0$, $Z_{\rm add}$ {\sl free}:  Here the initial gas
is assumed to be primordial and unenriched, while $Z_{\rm add}$ is determined
by the best fit to the observations.  This might be the case, for example,
if the bulge and halo began cluster formation at about the same time in
the early universe; the bulge region, sitting much deeper in the galaxy's
potential well, would have kept more of its gas and been able to
drive upward to higher metallicity.  (In practice, we adopt 
$Z_0({\rm phase 1}) \equiv 0$
for the halo stage in {\sl all} the model runs, since by hypothesis the
metal-poor cluster mode always starts from primordial 
or nearly-primordial gas.)

\noindent (b) $Z_0$ {\sl free}, $Z_{\rm add} = 0$:  Here the {\sl infalling}
gas in both phases is assumed to be primordial material, as is $Z_0$(phase 1).
Only $Z_0$(phase 2) is determined by the fit.  

\noindent (c) $Z_0 = Z_{\rm add}$:  In both stages we assume the infalling
gas from outside has the same abundance as the initial gas in the box.
Thus for the halo phase, $Z_0 = Z_{add} = 0$, while for the bulge phase,
both are determined in lockstep by the fitting procedure.

\noindent (d) $Z_0$ and $Z_{\rm add}$ free:  No assumptions are imposed
on either abundance, except for the usual $Z_0$(phase 1) = 0.

\section{Model Results and Discussion}

The primary results for our four starting assumptions are summarized
in Table 1.  In the last line of the table, mean uncertainties (corresponding
to one standard deviation) are listed.  Calculations of extensive model grids showed that the
initial masses $M_0$ and $M_{\rm add}(0)$ were 
were closely coupled and very weakly constrained.  (In some cases it was
possible to start with a virtually ``empty'' box and build the system
entirely from infalling material.)  The most 
interesting constraints were on $y_{\rm eff}$, $Z_{\rm add}$, and to a lesser
extent, the infall decay time $\tau$.  However, even these are partially
correlated:  for example, an increase in $Z_{\rm add}$ can be compensated by
a shorter time $\tau$, leading to almost the same final MDF.
In Table 1 we also list the typical (one-sigma) uncertainties of the fitted parameters.

All four initial assumptions are capable of leading
to good fits to the observed MDF; these are displayed in 
Figures 2(a-d).  Nevertheless,
a few clear trends emerge from inspection of the results.
For the ``halo'' stage, {\sl both the initial gas and infalling gas
need to be nearly unenriched, primordial material}, and the steep rise in
$dn/dZ$ at very low metallicity ($Z < 0.02 Z_{\odot}$) is the
characteristic mark of early heavy infall.  The effective
yield for the halo in all the models is always near $y_{\rm eff} \sim 0.0015$, 
five times smaller than the basic nucleosynthetic yield $y_0$.
This $y-$level suggests that $\sim$80\% or more of the protocluster gas in the halo 
was expelled before it could convert into stars, but the presence of infall
allows $y_{\rm eff}$ to be larger than in models
without inflow \citep{har76,bin98}.

For the ``bulge'' formation stage, $y_{\rm eff}$ is 
typically three times larger than for the halo phase.  If, as we expect, 
the mean difference is due mainly to
different ratios of ejected gas, then the bulge region, sitting much deeper
in the Galaxy's potential well, would have held onto more than half of 
its gas rather than only 20\%.

The exponential decay time for gas infall is consistently near $20 \delta t$
in phase 1, and 25 to 40 $\delta t$ in phase 2.
We cannot convert this to an actual
time interval without additional information on merging and star formation
rates which only more advanced models can provide.  
But if, following the GC
ages mentioned above, we assume that each
stage takes $\sim 2$ Gy over $\sim 100 \delta t$ steps, then in {\sl very}
rough terms the main period of heaviest gas infall 
should last about half a Gigayear in each stage.  

We emphasize here again that the assumption of {\sl no} infalling gas
($\tau \rightarrow 0$, corresponding to the closed-box model) does not 
reproduce the data adequately; such models produce too many clusters
in either phase at the low-metallicity end.  Said differently, a two-phase
{\sl closed-}box model, where the only free parameters are the yields
$y_{\rm eff}$, does not give the model enough freedom
to fit the data.  An early period of infall is both necessary and plausible.

\noindent {\sl Model (a):}
In this model even the metal-rich bulge clusters form starting from
initially unenriched gas.  However, the {\sl infalling} gas for stage 2 
must compensate by being substantially
pre-enriched (to about one-quarter Solar) to 
produce the correct secondary peak in the MDF.
For comparison, the hierarchical merging 
models \citep[see][]{col00,bea03} show that
the ``quiescent'' metal-poor mode starts with unenriched 
material but can continue upward
to near-Solar abundance as long as it is not interrupted.  The halo clusters
might have belonged only to the first part of this phase.

\noindent {\sl Model (b):}
In this model -- the poorest fit of the four -- the {\sl infalling} gas 
is always primordial, so $Z_0$ for stage 2
must start quite high to directly produce the 
secondary MDF peak (leaving the rather
unphysical spike at $Z = 0.34 Z_{\odot}$).  
Here, almost none of the bulge clusters fall in the 
metal-poor regime; there is almost no
overlap between the two parts of the MDF.

\noindent {\sl Model (c):}
This model starts with perhaps the most plausible 
assumption for both stages:  the infalling gas has
the same abundance as the initial $Z_0$.  The job 
of producing the secondary
MDF peak at the right place is then shared more 
or less equally between the initial
$Z-$level and the subsequent enrichment cycles 
from star formation.  The final maximum
enrichment level from stage 1 is fairly close 
to the initial $Z_0$ for stage 2, thus making it
easier for the halo clusters and field stars to 
form over similar time periods, unlike model (a) above.

\noindent {\sl Model (d):}
This model, which allows an unconstrained 
solution for both parametric abundances
$Z_0$ and $Z_{\rm add}$, ends up almost 
identical with model (a) and similar to (c) as well.

\section{Summary}

With an accreting-box chemical evolution model and some simple 
assumptions, we find that it is readily possible to describe the
MDF of the Milky Way globular clusters, with an accuracy and physical
basis superior to the standard double-Gaussian numerical fits
that pervade the literature.  
An unavoidable, and still rather arbitrary, key factor is that
we are forced to adopt two distinct phases of cluster formation
to produce the clear bimodality that the observations demand.

Within this context, however, a wide range of model parameters can
produce entirely adequate fits to the observations.  The combination
that we find most persuasive is that:

\noindent (a) The halo clusters formed from
near-primordial gas with a low yield $y_{\rm eff} \simeq 0.0015$
and shut down their formation at an early stage 
(this includes the possibility that
many of them formed in the small potential wells of 
dwarf satellites and were accreted later).

\noindent (b) The bulge clusters formed starting from mildly enriched gas
($\sim 0.2 Z_{\odot}$) and at $y_{\rm eff} \simeq 0.0045$.  In both
cases, a significant phase of early gas infall is necessary to reproduce
the observed MDF shape.

To the extent that two sharply distinct phases represent
something real in the actual history of the GCs
\citep[e.g.][]{san02}, this basic approach
needs to be explored further.  An important new piece
of the evolutionary puzzle now emerging is that the
MDFs of the globular clusters in several galaxies are        
strikingly different from the metallicity distributions of the halo 
and bulge {\sl field stars} \citep{hh01,dur01,har02}.  
In galaxies such as the LMC, M31, M32, and
NGC 5128 (and perhaps in giant E galaxies generally), the old-halo stellar
population is very broad and strongly weighted 
to moderately high metallicity near
[Fe/H] $\sim -0.5$.  These MDFs exhibit only a 
thin metal-poor tail and no trace of
the distinct bimodality characterizing the globular clusters.
This evidence reinforces the suspicion that the story of formation for the 
globular clusters had quite distinct elements.

Although we have discussed specific results only for the Milky Way, extensions
to other galaxies with clearly bimodal MDFs are obvious.  

\acknowledgments

This work was supported by the Natural Sciences and Engineering
Research Council of Canada through research grants to WEH.


\clearpage


\begin{deluxetable}{crcrrrcrrrrr}
\tablecaption{Accreting-Box Model Parameter Solutions \label{solutions}}
\tablewidth{0pt}
\tablehead{ &&& \multicolumn{4}{c}{Phase 1 Parameters} && 
\multicolumn{4}{c}{Phase 2 Parameters} \\ \cline{4-7}\cline{9-12}
\colhead{Model} & 
\colhead{$\chi_{\nu}$} & & 
\colhead{$y_{\rm eff}$\rule{0ex}{3ex}} &
\colhead{$\frac{Z_0}{Z_{\odot}}$} & 
\colhead{$\frac{Z_{\rm add}}{Z_{\odot}}$} & 
\colhead{$\frac{\tau}{\delta t}$} & &
\colhead{$y_{\rm eff}$} & 
\colhead{$\frac{Z_0}{Z_{\odot}}$} & 
\colhead{$\frac{Z_{\rm add}}{Z_{\odot}}$} & 
\colhead{$\frac{\tau}{\delta t}$}
}
\startdata 
$Z_0=0$ & 0.39 & ~~ & 0.0013 & 0 & 0.01 & 22 & ~~ & 0.0047 & 0 & 0.24 & 38 \\
$Z_{\rm add}=0$ & 0.90  & & 0.0016 & 0 & 0 & 18 & & 0.0046 & 0.34 & 0 & 27 \\
$Z_0=Z_{\rm add}$ & 0.38 & & 0.0015 & 0 & 0 & 22 & & 0.0047 & 0.25 & 0.25 & 38 \\
$Z_0, Z_{\rm add}$ free &  0.39 & & 0.0013 & 0 & 0.01 & 23  & & 0.0047 & 0.06 & 0.24 & 38 \\
\\
Uncertainty & & & 0.0002 & -- & -- &  3  & & 0.0004 & 0.06 & 0.03 & 4 \\
\enddata
\end{deluxetable}

\clearpage


\clearpage
\begin{figure}
\plotone{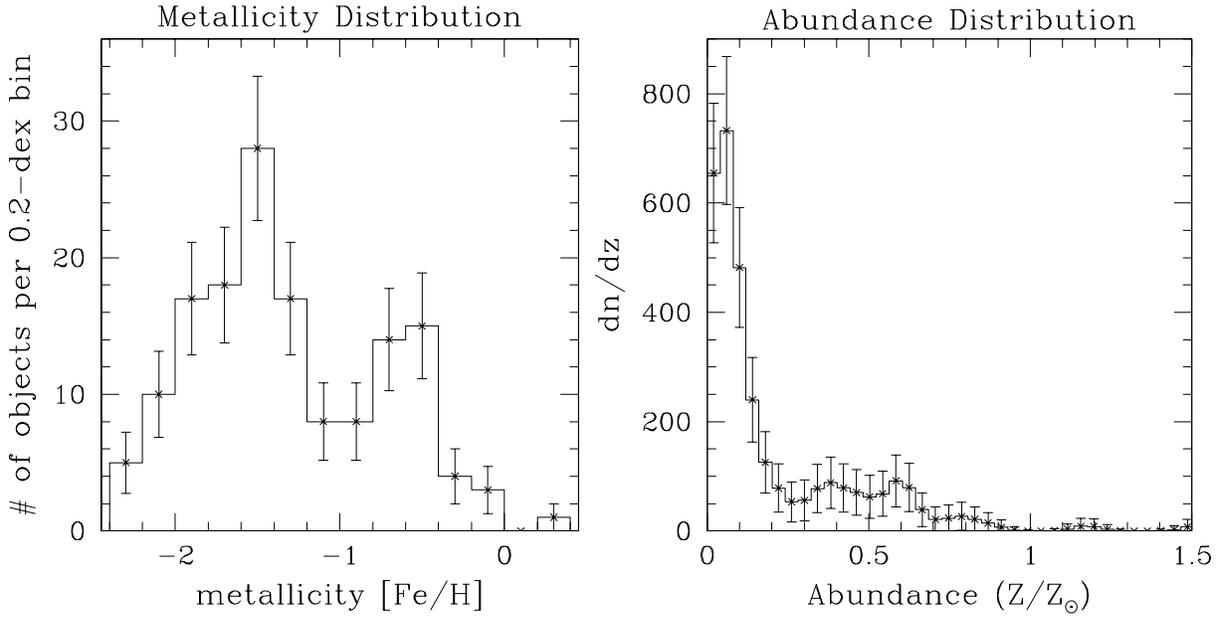}
\caption{Raw histograms showing the metallicity distribution function
for the globular clusters in the Milky Way, with data for 148 clusters from
the 2003 catalog edition of Harris (1996).  The left panel 
shows number of clusters per 0.2-dex
bin in [Fe/H], while the right panel shows the MDF in linear
form, as number per unit heavy-element abundance $Z/Z_{\odot}$.
Here log ($Z/Z_{\odot}$) = [Fe/H] + 0.3.  In the $(dn/dZ)$ graph the
bins have width $\Delta Z = 0.004$ and have been smoothed with a
kernel equal to the bin width.
\label{rawhisto}}
\end{figure}

\clearpage
\begin{figure}
\plotone{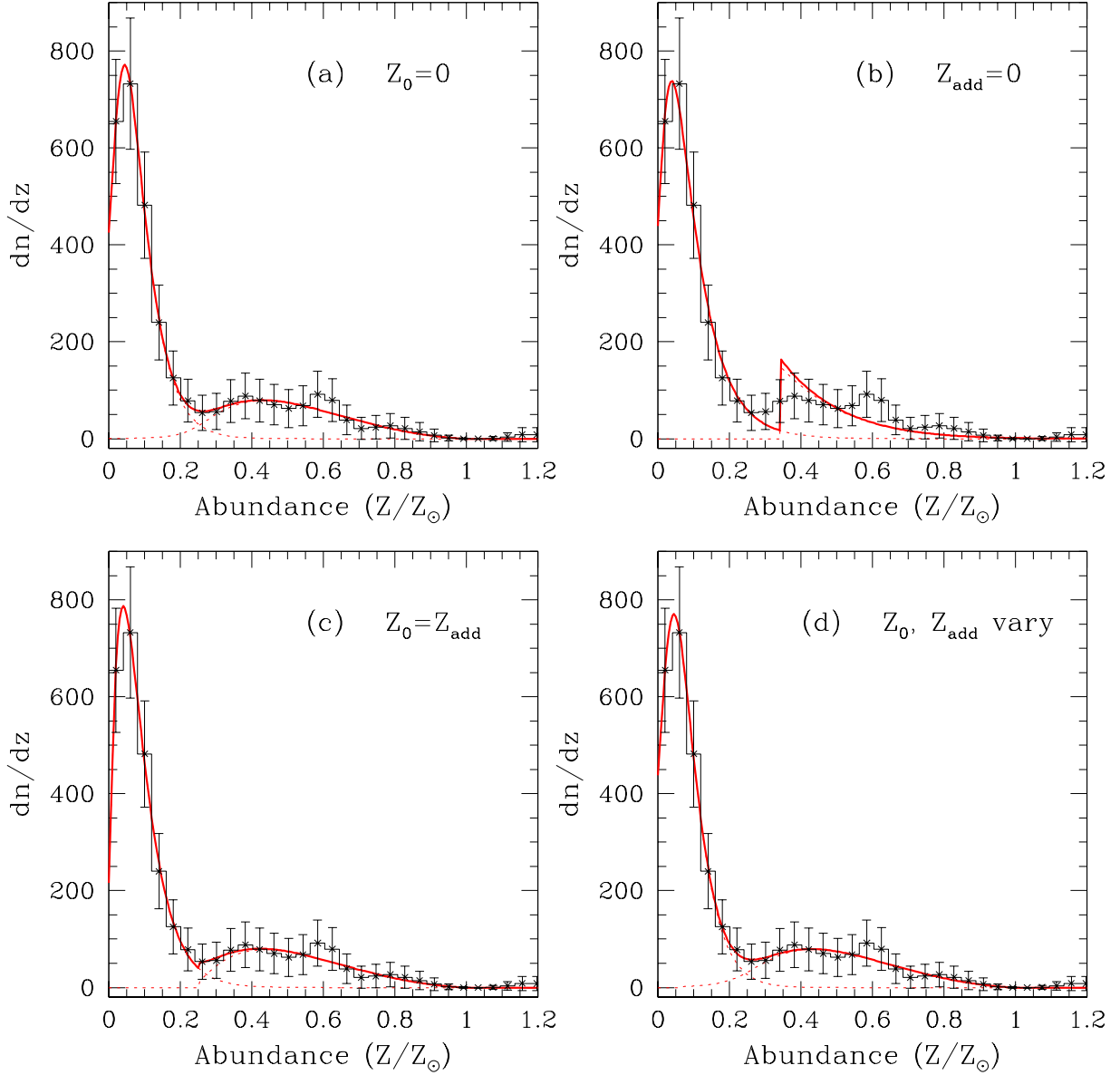}
\caption{Four models for the MDF of the Milky Way globular clusters.
The histogram with error bars in each panel represents the 
observed MDF for the Milky Way clusters in linear form ($dn/dZ$).
The heavy line is the best-fit model under
the assumptions described in the text,
while the dotted lines are the contributions from
the two individual phases.
The panels labeled a-d correspond to models a-d
respectively.
}
\label{modelfits}
\end{figure}


\clearpage

\begin{thebibliography}{}

\bibitem[Ashman \& Zepf(1992)]{ash92} Ashman, K.M., \& 
Zepf, S.E. 1992, \apj, 384, 50 
\bibitem[Beasley et al.(2002)]{bea02} Beasley, M.A., Baugh, C.M., 
Forbes, D.A., Sharples, R.M., \& Frenck, C.S. 2002, \mnras, 333, 383
\bibitem[Beasley et al.(2003)]{bea03} Beasley, M.A., Harris, W.E.,
Harris, G.L.H., \& Forbes, D.A. 2003, \mnras, 340, 341
\bibitem[Binney \& Merrifield(1998)]{bin98} Binney, J., \&
Merrifield, M. 1998, Galactic Astronomy (Princeton:  Princeton U. Press)
\bibitem[Chiappini, Matteuci, \& Gratton(1997)]{chi97} Chiappini, C., 
Matteuci, F., \& Gratton, R. 1997, \apj, 477, 765
\bibitem[Cole et al.(2000)]{col00} Cole, S., Lacey, C.G., Baugh, C.M.,
\& Frenk, C.S. 2000, \mnras, 319, 168
\bibitem[C\^ot\'e, Marzke, \& West(1998)]{cot98} C\^ot\'e, P., Marzke, R.O.,
\& West, M.J. 1998, \apj, 501, 554
\bibitem[C\^ot\'e, West, \& Marzke(2002)]{cot02} C\^ot\'e, P., West, M.J.,
\& Marzke, R.O. 2002, \apj, 567, 853
\bibitem[Durrell, Harris, \& Pritchet(2001)]{dur01} Durrell, P.R., Harris,
W.E., \& Pritchet, C.J. 2001, \aj, 121, 2557
\bibitem[Forbes, Brodie, \& Grillmair(1997)]{for97} Forbes, D.A.,
Brodie, J.P., \& Grillmair, C.J. 1997, \aj, 113, 1652
\bibitem[Forbes, Brodie, \& Larsen(2001)]{for01} Forbes, D.A., Brodie,
J.P., \& Larsen, S.S. 2001, \apjl, 556, L83
\bibitem[Gebhardt \& Kissler-Patig(1999)]{geb99} Gebhardt, K.,
\& Kissler-Patig, M. 1999, \aj, 118, 1526
\bibitem[Geisler et al.(1996)]{gei96} Geisler, D., Lee, M.G., \& 
Kim, E. 1996, \aj, 111, 1529
\bibitem[Harris(1996)]{har96} Harris, W.E. 1996, \aj, 112, 1487
\bibitem[Harris(2001)]{h01} Harris, W.E. 2001, in Star Clusters,
Saas-Fee Advanced Course 28 (New York:  Springer), ed.~L.Labhardt
\& B.Binggeli
\bibitem[Harris \& Harris(2001)]{hh01} Harris, W.E., \& Harris, G.L.H.
2001, \aj, 122, 3065 
\bibitem[Harris \& Harris(2002)]{har02} Harris, W.E., \& Harris, G.L.H.
2002, \aj, 123, 3108 
\bibitem[Harris, Harris, \& Geisler(2003)]{har03} Harris, W.E., Harris, 
G.L.H., \& Geisler, D. 2003, \aj, submitted
\bibitem[Hartwick(1976)]{har76} Hartwick, F.D.A. 1976, \apj, 209, 418
\bibitem[Kundu \& Whitmore(2001)]{kun01} Kundu, A., \& Whitmore, B.C.
2001, \aj, 121, 2950
\bibitem[Lacey \& Fall(1983)]{lac83}Lacey, C.G., \& Fall, S.M.
1983, \mnras, 204, 791
\bibitem[Larsen et al.(2001)]{lar01} Larsen, S.S., Brodie, J.P.,
Huchra, J.P., Forbes, D.A., \& Grillmair, C.J. 2001, \aj, 121, 2974
\bibitem[Larsen, Forbes, \& Brodie(2001)]{lfb01}Larsen, S.S., Forbes, D.A.,
\& Brodie, J.P. 2001, \mnras, 327, 1116
\bibitem[Neilsen \& Tsvetanov(1999)]{nei99} Neilsen, E.H.Jr., \&
Tsvetanov, Z.I.1999, \apj, 515, L13
\bibitem[Pagel \& Patchett(1975)]{pag75} Pagel, B.E.J., \& Patchett, B.E.
1975, \mnras, 172, 13
\bibitem[Prantzos(2003)]{pra03} Prantzos, N. 2003, \aap, 404, 211
\bibitem[Ryan \& Norris(1991)]{rya91} Ryan, S., \& Norris, J.1991,
\aj, 101, 1865
\bibitem[Santos(2002)]{san02} Santos, M.R. 2002, AAS Bull. 201, 108.05
\bibitem[Shetrone, C\^ot\'e, \& Sargent(2001)]{she01} Shetrone, M.D.,
C\^ot\'e, P., \& Sargent, W.L.W. 2001, \apj, 548, 592
\bibitem[VandenBerg(2000)]{van00} VandenBerg, D.A. 2000, \apjs, 129, 315
\bibitem[Whitmore et al.(1995)]{whi95} Whitmore, B.C., Sparks, W.B., Lucas, R.A.,
Macchetto, F.D., \& Biretta, J.A. 1995, \apjl, 454, L73
\bibitem[Zepf, Ashman, \& Geisler(1995)]{zep95} Zepf, S.E., Ashman, K.M.,
\& Geisler, D. 1995, \apj, 443, 570
\bibitem[Zinn(1985)]{zin85} Zinn, R. 1985, \apj, 293, 424

\end{thebibliography}
\end{document}